\documentclass[runningheads]{llncs}
\usepackage{graphicx}
\usepackage{subfigure}
\usepackage{hyperref}
\usepackage[dvipsnames]{xcolor}

\begin{document}
\title{An Overview of Human Activity Recognition Using Wearable Sensors: Healthcare and Artificial Intelligence}
%
%

\author{Rex Liu\inst{*}\inst{1}\and
Albara Ah Ramli\inst{*}\inst{1}\and 
Huanle Zhang\inst{1}\and 
Erik Henricson\inst{2}\and 
Xin Liu\inst{1}}

\renewcommand{\thefootnote}{\alph{footnote}}
\footnotetext[0]{* These two authors contributed equally to this work.}
\authorrunning{}
\titlerunning{An Overview of Human Activity Recognition Using Wearable Sensors}
%

\institute{
Department of Computer Science, School of Engineering\\
\and Department of Physical Medicine and Rehabilitation, School of Medicine\\
University of California, Davis\\
\email{\{rexliu,arramli,dtczhang,ehenricson,xinliu\}@ucdavis.edu} \\
}

\maketitle              

\begin{abstract}
With the rapid development of the internet of things (IoT) and artificial intelligence (AI) technologies, human activity recognition (HAR) has been applied in a variety of domains such as security and surveillance, human-robot interaction, and entertainment. Even though a number of surveys and review papers have been published, there is a lack of HAR overview papers focusing on healthcare applications that use wearable sensors. Therefore, we fill in the gap by presenting this overview paper. In particular, we present our projects to illustrate the system design of HAR applications for healthcare. Our projects include early mobility identification of human activities for intensive care unit (ICU) patients and gait analysis of Duchenne muscular dystrophy (DMD) patients. We cover essential components of designing HAR systems including sensor factors (e.g., type, number, and placement location), AI model selection (e.g., classical machine learning models versus deep learning models), and feature engineering. In addition, we highlight the challenges of such healthcare-oriented HAR systems and propose several research opportunities for both the medical and the computer science community. 


\end{abstract}
\section{Introduction} 

Human activity recognition has been actively researched in the past decade, thanks to the increasing number of deployed smart devices such as smartphones and IoT devices. Based on the type of data being processed, a HAR system can be classified into vision-based and sensor-based. This paper targets wearable-sensor HAR systems in healthcare, which are the most prevalent type of sensor-based HAR systems~\cite{survey20pattern}. More importantly, wearable-sensor HAR systems do not suffer from severe privacy issues like vision-based HAR systems, making wearable-sensor HAR systems suitable for healthcare applications.  
In a wearable-sensor HAR system, a user wears portable mobile devices that have built-in sensors. The user's activities can then be classified by measuring and characterizing sensor signals when the user is conducting daily activities. 

HAR for healthcare has many potential use cases, including (1) Moving gait diagnosis from expensive motion labs to the community. Gait analysis can be used in many healthcare applications, such as stroke detection, gait modification (to prevent failing), and certain disease early detection. (2) Cognitive behavior monitoring and intervention for children and adults with attention-deficit/hyperactivity disorder (ADHD). We can leverage sensors to investigate whether fidgeting positively or negatively affects attention. (3) Stroke-patient hospital direction. When a patient is in an ambulance, a life-and-death question is whether the patient has extensive brain hemorrhage. If so, the patient should be directed to a hospital that can treat such cases. UCSF has developed a device based on an accelerometer sensor to help make this critical decision. (4) Epilepsy and Parkinson's disease study. Doctors have collected a significant amount of data on electrophysiology and episodic memory in rodents and human patients. The analysis of such sensing data can be used for various disease identification and treatment purpose. (5) An expensive device, called Vision RT, is used to ensure radiation therapy is delivered safely to cancer patients (due to patient motion). It is worth exploiting sensors to detect the patient's movement while taking radiation therapy for the less affluent communities. 

However, building practical wearable-sensor HAR systems for healthcare applications not only has challenges (e.g., sensor setup, data collection, and AI model selection) that are faced by traditional wearable-HAR systems, but also challenges that are unique to the healthcare domain. For example, in addition to the overall AI model accuracy (averaging results of all users), clinicians are concerned about the model stability (i.e., the model has approximately the same accuracy for each user) and model interpretability (e.g., to discover patient movement patterns that are specific to some symptoms).



Therefore, we present this overview paper in the hope to shed light on designing wearable-sensor HAR systems for healthcare applications. To illustrate the system considerations, we share two of our healthcare systems: one for identifying the early mobility activities of ICU patients~\cite{ICUpaper} and the other one for the gait analysis of DMD patients~\cite{DMD_PAPER}. Our projects demonstrate that HAR systems for healthcare not only have commonalities such as data processing pipelines but also differences in terms of sensor setup and system requirements.

We organize this paper as follows. First, we provide the preliminaries of HAR systems. Next, we introduce our HAR systems for ICU patients and DMD patients. Then, we explain the considerations when designing a HAR system. Last, we highlight the challenges of applying wearable-sensor-based HAR systems to healthcare, and propose several research opportunities. Last, we conclude this paper. 
\section{Human Activity Recognition: A Primer}

\begin{figure}[!t]
\centering
\centerline{
\includegraphics[width=4in]{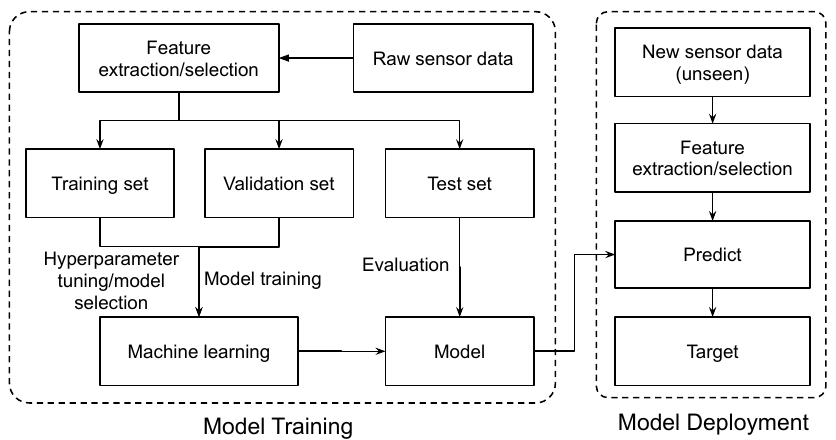}}
\caption{
General data flow for the two-stages of HAR systems: model training and model deployment. 
}
\label{fig:data_flow}
\end{figure}

Given the short time-length data of wearable sensors, a HAR system needs to recognize the activity from which the data is generated. Thanks to the rapid advancement of AI technology, AI algorithms/models are increasingly adopted for recognizing the activity from the sensor data. Figure \ref{fig:data_flow} illustrates the general data flow for an AI-based HAR system, which can be divided into two stages: model training and model deployment.

In the model training stage, an AI model is trained and tailored for the specific application. To achieve an accurate AI model, the following steps are often applied. First, raw sensor data from different activities should be collected. The quality of collected data significantly affects the AI model performance. The collected data is required to be diverse, representative, and large in the number of samples. Afterward, the raw data is divided into fixed-length or dynamic-length segments (i.e., time windows) \cite{window16pervasive}. Then, feature extraction is used to extract potentially useful features from the data segmentation, and feature selection is adopted to remove irrelevant features~\cite{feature14sic}. To alleviate the overfitting problem of the trained model, the set of processed features are divided into a training set, a validation set, and a test set. During the AI model training, we use the training set to tune the AI model and the validation set to measure the model's accuracy. After we finish the model training, we use the test set to evaluate the trained model. The trained model is deployed to real-world applications if its accuracy is satisfactory. Otherwise, the whole model training stage is performed repetitively by exploring different configurations, such as applying other feature extraction methods and changing AI models. 

In the model deployment stage, the same data processing (e.g., segmentation, feature extraction, and selection) is applied to the new and unseen sensor data, and the trained model is executed on the processed data. 
It is possible that the trained model may not work as expected in a real deployment, probably due to the model over-fitting or the lack of generality in the collected dataset \cite{deep19prl}. In this situation, the system designer needs to revert to the model training stage.

\section{HAR Applications in Healthcare}
\label{sect:application}

\begin{figure}[!t]
\centering
\subfigure[ICU-patient]{%
\label{fig:device_setup_icu}
\begin{minipage}[b]{1.5in}
\qquad \qquad
\includegraphics[height=1.3in]{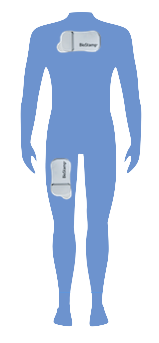}
\end{minipage}
}%
\subfigure[DMD patient]{%
 \label{fig:device_setup_dmd}
 \begin{minipage}[b]{1.5in}
 \qquad \quad
 \includegraphics[height=1.3in]{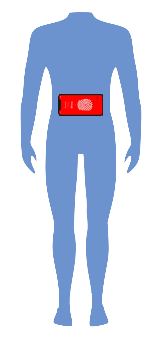}
 \end{minipage}
}%
\caption{Device setups in our HAR projects. (a) We use two accelerometer devices to recognize the early mobility activities of ICU patients. One device is on the chest and the other device is on the thigh. (b) We use one smartphone that captures accelerometer data to identify DMD patients. The phone is located at the backside body.}
\label{fig:device_setup}
\end{figure}

Clinicians have already applied wearable sensor-based HAR systems in healthcare, thanks to the development of more lightweight wearable devices, greater computation capability, and higher accurate AI algorithms. This section presents our two HAR healthcare projects to illustrate the considerations when designing HAR systems for healthcare applications with different goals.

\subsection{Case 1: Identification of Early Mobility Activity for ICU Patients}

Due to long periods of inactivity and immobilization, patients become weak when recovering from major illnesses in ICU~\cite{icu15plos}. If ICU patients' activities can be accurately recognized, clinicians can provide an optimal personalized dose of mobilities for ICU patients' different illness conditions.  Therefore, doctors and researchers are extremely interested in ICU patients' early mobilization, which is an effective and safe intervention to improve functional outcomes~\cite{ema12cptj}. However, early mobility activity (EMA) research is limited by the lack of accurate, effective, and comprehensive methods to recognize patients' activities in ICU.

\begin{figure*}[!t]
\centering
\centerline{
\subfigure[ICU patient]{%
\label{fig:raw_data_icu}
\includegraphics[height=1.0in]{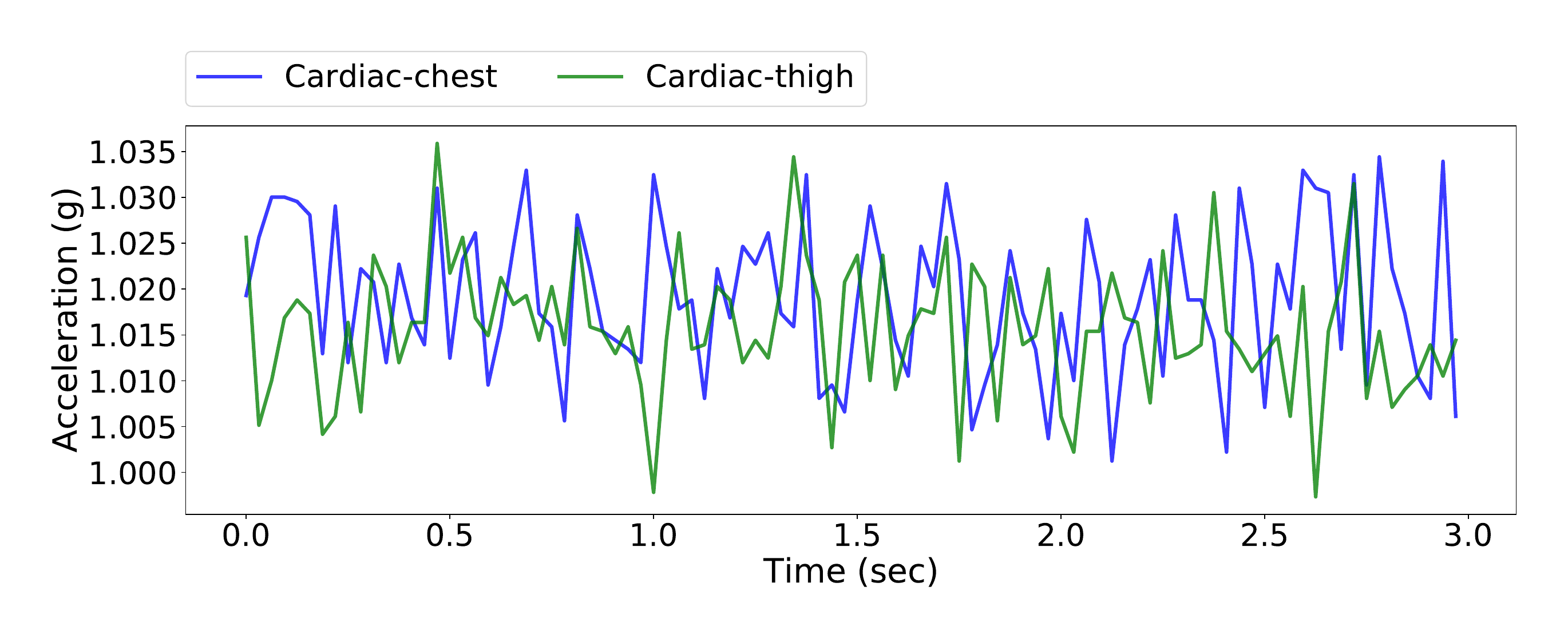}}%
\subfigure[DMD patient]{%
\label{fig:dmd_vs_control}
\includegraphics[height=1.0in]{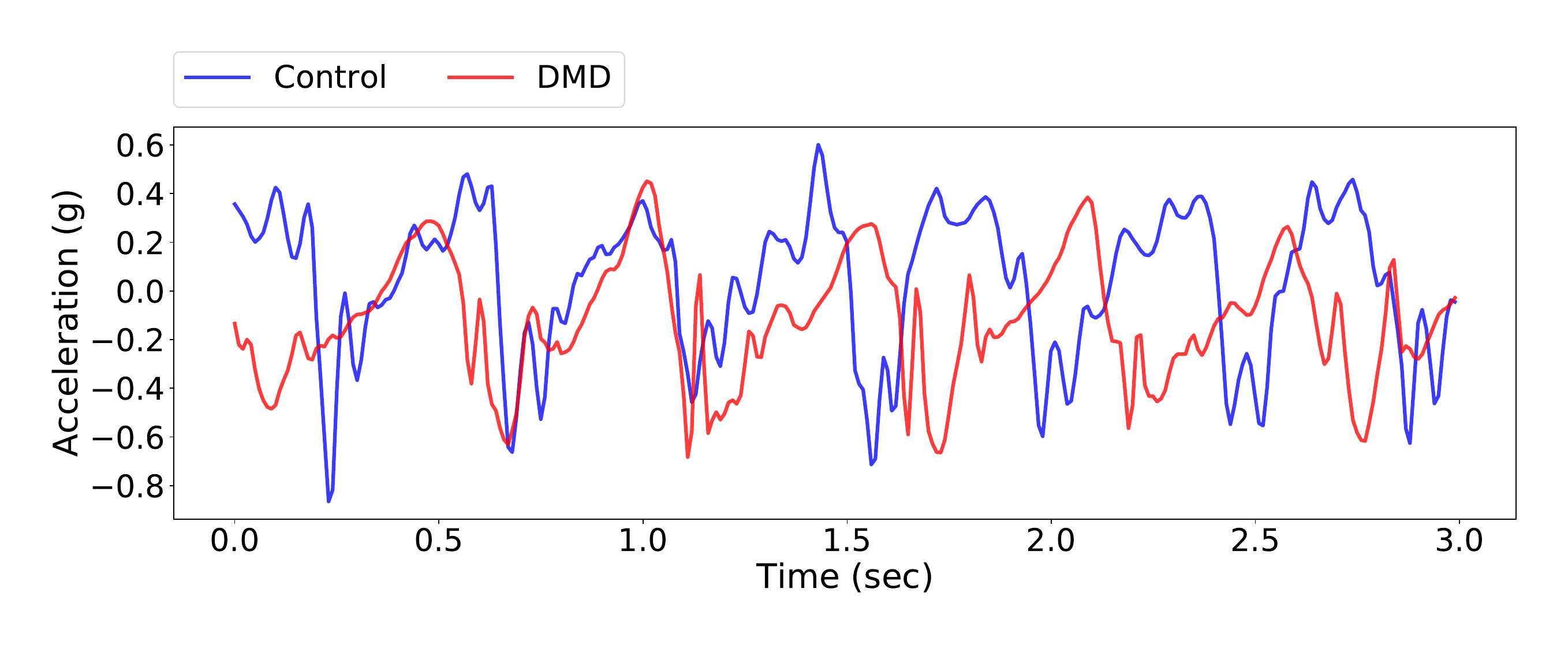}}%
}
\caption{Illustration of accelerometer data in our projects. (a) The z-axis of the accelerometer data from the two on-body devices when an ICU patient is performing the cardiac activity. (b) The z-axis of the accelerometer data, which shows the difference in gait characteristics between a DMD patient and a healthy person.}
\end{figure*}

We propose a wearable sensor-based HAR system for recognizing the EMA of ICU patients~\cite{ICUpaper}. In our system, Each ICU patient wears two accelerometer devices: one on the chest and the other on the thigh, as shown in Figure \ref{fig:device_setup_icu}. Each device continuously collects 3-axis accelerometer data at a sampling rate of 32 Hz. Figure \ref{fig:raw_data_icu} plots the accelerometer data when an ICU patient sits on the cardiac chair to achieve an optimal resting position. This project aims to classify 20 types of ICU-related activities (e.g., reposition, percussion). 

This project has two main challenges in designing the HAR system for ICU patients. (1) Label Noise. Because the time lengths for accomplishing an early mobility activity are different for ICU patients with varying health conditions, it is laborious and time-consuming work for clinicians to annotate sensor data for each second in the real world. Therefore,  our EMA sensor data are annotated for each minute by a medical expert after data collection. However, one-minute length is exceedingly long for some early mobility activities such as Reposition, which the patient needs less than 20 seconds to accomplish. This annotation process introduces the label noise in our EMA dataset, which decreases the accuracy of the model. (2) Sensor Orientation. In the actual data collection process and possible future applications, we cannot guarantee that the orientations of all accelerometers are the same, and different orientations of the accelerometers lead to different meanings of XYZ coordinate values. Therefore, without careful feature extraction and selection, the AI model generalizes poorly to different patients, affecting the system performance in practice.

To tackle these challenges and improve the accuracy of recognizing ICU patient's activities, we explore the following techniques. (1) We propose a segment voting process to handle the label noise. Specifically, each one-minute sensor data is divided into multiple fixed half-overlapped sliding segments (time windows). We train our AI model using the segments. To predict each one-minute sensor data activity,  we apply our trained model to each segment. The final prediction result for the one-minute data is the activity that has the majority vote among the prediction of all segments. Our segmenting method improves the model accuracy by $\sim$4.08\% and reduces the model instability by $\sim$9.77\%~\cite{ICUpaper}. Our experiments also demonstrate that the number of sensors contributes to eliminating label noise in our dataset. As shown in Figure \ref{fig:raw_data_icu}, the increase in the number of sensors conveys more information, and thus improves the system's accuracy. (2) We identify and extract features that are not sensitive to sensor orientations to tackle the sensor orientation problem. Our features improve both the accuracy and the stability of AI models compared to the model trained on commonly used features.

\subsection{Case 2: Identification of Gait Characteristics for DMD Patients}

Duchenne muscular dystrophy (DMD) is a genetic disorder disease that affects the dystrophin protein, essential for keeping muscle cells intact. It has an estimated incidence of 1:5000 male births, and untreated boys become wheelchair-bound by the age of 12 years and die in their late teens to early 20s~\cite{dmd15jpch}. There is presently no cure for DMD disease. Nonetheless, gene repair interventions and other preventive therapies can be initiated as early as possible to slow the disease's progress and prevent secondary conditions. Therefore, it is important to identify children with DMD early in the course of their disease and have tools for quantitative evaluation of their gait in both the clinic and community environments.

We designed a wearable sensor-based HAR system to identify gait characteristics associated with the progression of gait abnormalities in children with DMD and to differentiate those patterns from those of typically developing peers~\cite{DMD_PAPER}~\cite{dmdpropose20davis}
To leverage this idea, we design a HAR system in which we use a smartphone to capture accelerometer data from the participants. As Figure \ref{fig:device_setup_dmd} illustrates, participants wear a smartphone at the back of the hips over the spine (lumbosacral junction) at a location that is the closest surface point to the body’s center of mass. Each smartphone collects 3-axis accelerometer data at a sampling rate of 30 Hz with the same phone orientation. 

We recruited ambulatory participants with DMD between 3 and 13 years of age and typically developing controls of similar ages. We ask participants to perform exercises at various times, speeds, and distances such as free walk and 6-minute walk, as specified by the north star ambulatory assessment (NSAA) standard~\cite{nsaawebsite}. Figure \ref{fig:dmd_vs_control} shows the gait pattern difference between a DMD patient and a healthy person when they are walking. 

We found that classical machine learning and deep learning, after hyper-parameter fine-tuning and cross-validation on seven different gait activities, led to the best performance with an accuracy exceeding 91\% on the 6-min-walk-test activity~\cite{DMD_PAPER}. We demonstrate that by using AI techniques and an accelerometer, we can distinguish between the DMD gait and typically developing peers.

There are two main challenges in designing our HAR system for the DMD application: clinical interpretability and data sparsity. (1) Clinical Interpretability. Medical practitioners desire not only a high prediction accuracy but also an interpretation of the prediction result. (2) Data Sparsity. In the healthcare domain, collecting diverse and sufficient data is challenging, especially for fatal diseases such as DMD.

We explore the following techniques to tackle these challenges. (1) To interpret AI model outcomes, we plan to link the clinical measurements with the model's extracted features by leveraging advanced AI models such as interpretable CNN~\cite{cnn18cvpr}.
However, it is an active, challenging task to find which clinical measurements correlated with the AI model features, especially for deep learning models. (2) To overcome the lack of data, we plan to use Generative Adversarial Network (GAN)~\cite{gan} or synthetic minority over-sampling technique (SMOTE)~\cite{smote02jair} to generate more data samples. 

\subsection{Summary of Our Projects}

Our two projects target different healthcare applications with different goals:  recognizing ICU patients' activities and distinguishing DMD gait patterns from those typically developing controls. The ICU project focuses on the system performance to assist the doctor in better understanding patients' recovery. While achieving high system performance, the DMD project interprets the model results further and discovers disease-specific patterns to determine the patient's condition and progression. Our example projects demonstrate the effectiveness and potential of wearable sensor-based HAR systems in healthcare. However, due to the different goals, different healthcare applications may have additional HAR system considerations. For example, our two projects adopt a different number of devices (2 versus 1) and device position (chest and thigh versus central mass body). In addition, our projects also apply different feature extractions (time and frequency domain versus clinical). In the next section, we present design considerations for building HAR systems.

\section{System Design}
\label{sect:design}

This section covers three design considerations essential for HAR systems, i.e., sensor, feature extraction and selection, and AI model selection.

\subsection{Sensor}

Sensors play an essential role in wearable HAR systems. Different HAR systems adopt various sensor configurations regarding the type of sensors, the sensor position and orientation, and the number of sensors.

\subsubsection{Sensor Types}

There are several types of sensors. Each sensor captures a different raw movement signal. The most commonly-used wearable sensors in HAR systems are accelerometer, gyroscope, and electrocardiography (ECG). The accelerometer sensor captures the acceleration signal that is useful for recognizing movements such as walking, running, and jumping. Gyroscopes capture the rotation movements used commonly in recognizing swinging, turning, and repositioning. ECG captures the heart rate and rhythm, which helps distinguish between intensive and light exercises.

However, many activities include both directional and rotational movements. Therefore, using one sensor type is not adequate. As a result, multiple types of sensors (e.g., accelerometer and gyroscope) are used in various application scenarios to maximize accuracy. However, using multiple types of sensors is challenging due to the increased complexity of the system in terms of synchronization issues \cite{sync13communication}.

\subsubsection{Sensor Position and Orientation}

Different positions and orientations of devices affect the data features and thus the model accuracy in predicting different activities \cite{position06bsn}. However, there have not yet been systematic comparisons of the number, type, and location of sensors to determine whether an optimal array design can capture data across a wide range of human activities and disease states. In many cases, the device position and orientation are decided by the empirical experience of clinicians.

\subsubsection{Number of Sensors}

Generally, a large number of sensors require demanding storage and computation capability. On the other hand, more sensors can collect more diverse data, which is beneficial for improving model performance~\cite{num02swc}. Therefore, to decide the optimal number of sensors, researchers need to carefully consider many factors such as cost, power consumption, and accuracy target as well as the feasibility of long-term use in the community to collect real-world information~\cite{jarchi2018review}.

\subsection{Feature Extraction and Selection}

In addition to the hardware setup, feature extraction and selection significantly affect the overall system performance. Before applying feature engineering to the data, the input data needs to be segmented. 

\subsubsection{Data Segmentation}

HAR systems collect data constantly via wearable sensors to identify possible activities. Data segmentation is applied to divide comparatively long time data into short fragments (time windows) that are suitable for AI models to learn. There are two types of data segmentation: fixed-length and dynamic-length~\cite{window16pervasive}. For fixed-length segmentation, if the time window is too short, the extracted features from the fragments are insufficient to capture the activity; on the other hand, if the time window is too long, a fragment is likely to contain multiple activities. The system accuracy deteriorates in both cases. In comparison, a dynamic-length data segmentation adopts an adaptive length of fragments corresponding to the characteristics of input data. Ideally, dynamic data segmentation generates fragments, in which each fragment only contains a single and complete activity. However, dynamic data segmentation is much more complex than fixed data segmentation, and thus are not as widely adopted by existing works as fixed-length segmentation. 

\subsubsection{Feature Extraction} 

Feature extraction is then applied to extract important features from the data fragments~\cite{feature14sic}. It can be broadly classified into time-domain and frequency-domain methods. In time-domain feature extraction, metrics such as median, variance, mean, and skewness are calculated over the amplitude variations of data over time. Time-domain features are lightweight to compute and thus are friendly to low-profile embedded devices and real-time applications. In comparison, frequency-domain features calculate the frequency variations of data over time. They include metrics such as spectral entropy, spectral power, and peak frequency. The computation overhead of frequency-domain features is generally much greater than time-domain features. In reality, most existing HAR systems adopt both time-domain features and frequency-domain features, in the consideration of the tradeoff among factors such as system accuracy, computation overhead, and power consumption. 

\subsubsection{Feature Selection} 

Feature selection is often adopted in order to reduce system complexity. It measures the importance of features and then removes irrelevant features. Feature selection is roughly divided into three methods: filter methods, wrapper methods, and embedded/hybrid methods~\cite{feature14sic}. Filter methods select a subset of features by exploiting inherent characteristics of features, whereas wrapper methods use classifiers to estimate the useful features.  On the other hand, the embedded/hybrid methods combine the results from filter methods and wrapper methods~\cite{survey20pattern}. By carefully selecting features, the AI model accuracy can be significantly improved. However, in healthcare HAR systems, pursuing high accuracy is not the sole goal, as the features are often manually decided by medical experts for identifying patients. Therefore, healthcare HAR systems require feature extraction and selection that is meaningful for clinicians and meanwhile achieves high prediction accuracy.

\begin{figure*}[!t] \centerline{
\includegraphics[width=5.0in]{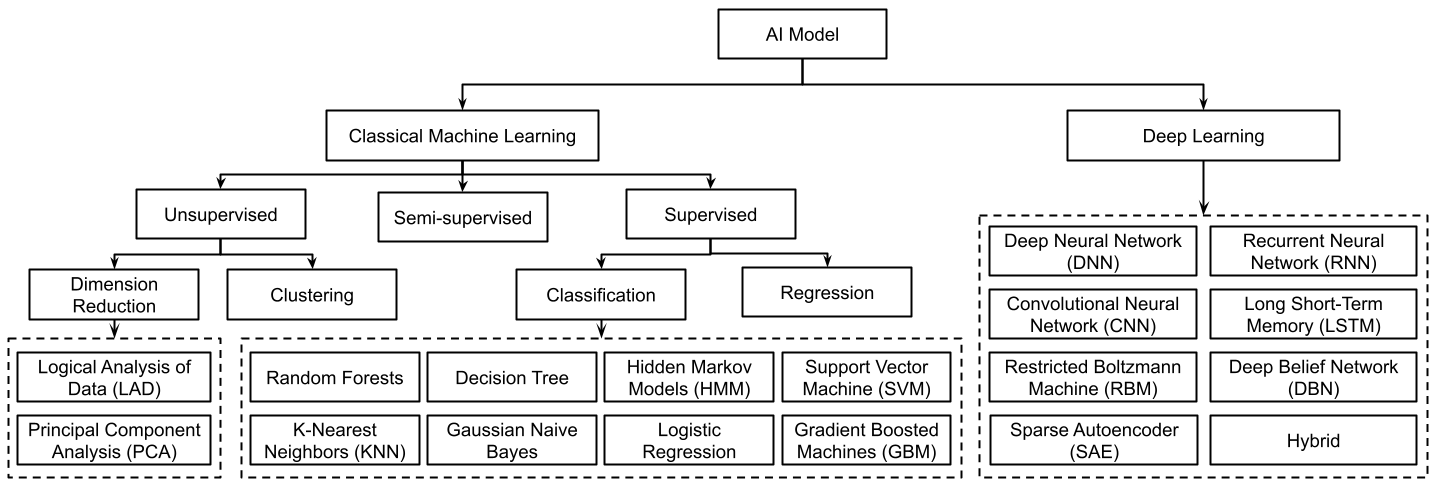}}
\caption{Classical machine learning and deep learning algorithms used in HAR systems.
}
\label{fig:ai_model}
\end{figure*}

\subsection{AI Model Selection}
In the HAR field, classical machine learning algorithms and deep learning algorithms have been explored and applied, which is summarized in Figure \ref{fig:ai_model}. Both classical machine learning algorithms and deep learning algorithms have different advantages and disadvantages. 


\textbf{Dataset requirement and system running overhead.}
The data collection process in the healthcare scenario is challenging because of the severe privacy issue and rare incidence rate of some medical activities. Therefore, in most healthcare applications, the database size is small. Correspondingly, classical machine learning models are more preferred because they work well with medium-size datasets. In contrast, even though deep learning models achieve better accuracy, they usually require a large amount of data for training. 
Real-time performance is another critical factor for some healthcare applications \cite{bwcnn}. For example, \cite{realtime_cranial} uses cranial accelerometers to detect stroke in an ambulance to decide whether to send the patient to a specialist stroke hospital for special treatment. Therefore, lightweight models are preferred in this use case.
In addition to the running overhead of the AI models, the processing time of feature extraction also affects the model selection, because different model structures adapt differently to the extracted features. 

\textbf{System interpretability. }
The features extracted from the sensor data are helpful to understand the pattern of some specific diseases to find out the pathological characteristics of the disease. For example, we extract the temporal/spatial gait characteristics from sensor data to evaluate the gait changes associated with DMD. Classical machine learning models are easier to interpret the model's decision, especially in decision tree models. Even though there is a great deal of work in interpreting deep learning models, deep learning models have the reputation of poor interpretability.

\section{Challenges and Opportunities}

Wearable sensor-based HAR systems are promising for a variety of healthcare problems. However, there are several challenges in fully exploiting them to build satisfactory HAR systems for healthcare. 
In this section, we identify challenges as well as research opportunities of HAR systems for healthcare.

\subsection{Data Sparsity}
The most commonly used algorithms for the HAR system in healthcare are the supervised learning algorithms that need extensive labeled data. For some daily living activities such as walking and running, researchers could get a significant amount of the labeled data from the public dataset or the raw sensor data collected and annotated by themselves. However, for some specific human activities related to healthcare, such as the therapeutic activities of patients, researchers could not get enough sensor data since these activities are low-probability events compared with daily life activities. Furthermore, it also takes time and effort to locate the sensor data of these specific activities from the daily data and label them. For example, when patients recover from surgery, they need some range of motion(ROM) exercises several times a day to make their joints and muscles flexible and strong again. Because of the fixed and limited collection times per day and the limited number of patients are involved, raw sensor data for ROM becomes insufficient, affecting the HAR system's performance. Therefore, building HAR systems with high accuracy on small datasets in healthcare is one of the most significant challenges.

Meta-learning is one of the approaches to solve this challenge. Meta-learning aims to optimize models which can learn efficiently in a small dataset when dealing with new categories.  In ~\cite{meta-learning}, researchers present a meta-learning methodology based on the Model-Agnostic Meta-Learning algorithm~\cite{maml} to build personal HAR models. In ~\cite{few-shot}, researchers use few-shot learning to transfer information from existing activity recognition models. However, it is unclear whether these techniques work well for medical applications. So more research is needed to explore the feasibility of transferring knowledge from daily living activities to specific activities related to healthcare.

\subsection{Model Interpretability}
In HAR applications in healthcare, an increasing number of applications focus on the interpretability of the model to extract relevant features, in order to describe the severity of the disease and track the progression of the disease~\cite{DMD_PAPER}. In addition, notwithstanding the benefit of deep learning in HAR, the underlying mechanics of machine learning are still unclear. So, various studies are trying to explain the deep learning model for the recognition of human activities. The common approach to interpreting the deep learning model is to compute the importance of each part of the input. In ~\cite{interpret-position}, researchers propose an interpretable convolutional neural network to select the most important sensor position for some specific activities. Instead of computing the importance of each part of the input, another approach is to make a sequence of selections about which part of the input is essential for the model training~\cite{deep-rl}. More research is required to adopt these methods to HAR systems for healthcare.


\subsection{Concurrent Activities}
Most of the existing HAR research focuses on single-labeled activity, recognizing only one activity of the given data segment. However, in real-world healthcare scenarios, humans can perform multiple activities concurrently. For example, patients can do ROM exercises and percussion therapy at the same time. The AI model performance deteriorates for concurrent activities. On the other hand, designing models to recognize multiple activities per data segment is a challenging task.

\subsection{Composite Activities}

In healthcare applications, optimizing HAR algorithms to identify composite activities in the community is ultimately more desirable than recognizing a single type of task. For example, when a patient moves from bed to the chair, the patient performs various activities, including sitting from supine in the bed, pivoting to place feet on the floor, standing from sitting, walking a few steps, and then sitting down on a chair. Therefore, it is preferred that an AI model can directly recognize the composite activity. 

\subsection{Privacy}
Wearable sensor-based HAR systems do not suffer from severe privacy issues as camera-based vision systems. However, since HAR applications continuously capture user data and recognize user activities, they may leak users' personal information if data are not secured. Therefore, secure data sharing and safe data storage are imperative for healthcare applications. To alleviate sensitive information during model training,  adversarial loss functions are leveraged to guard against privacy leakage~\cite{security17ijcai}. 
In addition, federated learning is a promising solution, which trains a global model without exposing local devices' private data~\cite{fl17aistats}.

\subsection{Opportunities of HAR for Healthcare}

Through our experience with HAR systems for healthcare, we identify the following research opportunities.
\begin{itemize}
    \item \textbf{Community-based healthcare}. Community-based healthcare requires that user devices are lightweight and affordable for the public. In addition, instructing the non-expert users/patients should be straightforward to follow. We can use digital sensing capability and the popularity of mobile devices to enable large community-based prescreening for various diseases and early signs of diseases. This can be done in a privacy-preserving manner in the sense that data does not need to leave a local device if necessary. For example, our DMD project enables community-based diagnosis during the pandemic and in rural areas where specialty labs are hundreds of miles away.
    \item \textbf{Chronic disease prevention and intervention.}  For chronic diseases, it is essential to capture the behaviors of patients in the long run. To this end, gait analysis, motion monitoring, ECG, and other vital signals (such as continuous glucose monitoring) can play a key role. 
    \item \textbf{Health aging}. With the decreased fertility rates and the increased life expectancy, population aging is becoming common for most countries. Therefore, building HAR systems for healthy aging is beneficial for senior citizens and society as a whole. We anticipate that gait and motion monitoring and diagnosis will play a critical role in healthy aging.
\end{itemize}

\section{Conclusion}

It is gaining popularity by applying wearable sensors to recognize and analyze human activities for the healthcare domain. For example, we leverage HAR systems to recognizing patients' early mobility activities in ICU and to analyzing the symptoms of DMD patients. This overview paper covers the system design of HAR systems based on wearable sensors, focusing on healthcare applications. We emphasize the essential components of HAR systems, including sensor factors, data segmentation, feature extraction and selection, and AI model comparison. We also highlight the challenges and opportunities of HAR systems for healthcare. 

\section{Acknowledgement}
The authors thank Esha Datta for their  contribution to the manuscript. 
%
%
%
%
\bibliographystyle{unsrt}
\bibliography{reference} 

\end{document}